\documentclass[aps,twocolumn,pre,showpacs,color,psfig,epsf]{revtex4}
\usepackage{amssymb,amsfonts,amsmath}
\usepackage{color}
\usepackage{amsfonts}
\usepackage{epsf}
\usepackage{graphicx}
\usepackage{xcolor}

\newcommand\zed{{\bar \zeta}}

\begin{document}

\title{Spontaneous symmetry breaking in active droplets provides a generic route to motility}

\author{E. Tjhung, D. Marenduzzo, M. E. Cates} 
\affiliation{SUPA, School of Physics and Astronomy, University of
Edinburgh, Mayfield Road, Edinburgh EH9 3JZ, UK}

\begin{abstract} 
We explore a generic mechanism whereby a droplet of active matter acquires motility by the spontaneous breakdown of a discrete symmetry. The model we study offers a simple representation of a ``cell extract" comprising, e.g., a droplet of actomyosin solution. (Such extracts are used experimentally to model the cytoskeleton.) Actomyosin is an active gel whose polarity describes the mean sense of alignment of actin fibres. In the absence of polymerization and depolymerization processes (`treadmilling'), the gel's dynamics arises solely from the contractile motion of myosin motors; this should be unchanged when polarity is inverted. Our results suggest that motility can arise in the absence of treadmilling, by spontaneous symmetry breaking (SSB) of polarity inversion symmetry. Adapting our model to wall-bound cells in two dimensions, we find that as wall friction is reduced, treadmilling-induced motility falls but SSB-mediated motility rises. The latter might therefore be crucial in three dimensions where frictional forces are likely to be modest.
At a supra-cellular level, the same generic mechanism can impart motility to aggregates of non-motile but active bacteria; we show that SSB in this (extensile) case leads generically to rotational as well as translational motion. 

\end{abstract}

\maketitle

Living cells can move themselves around in a variety of different conditions and environments, and they exploit a range of strategies and mechanisms to do so. 
Uncovering the generic pathways to cell motility remains central to many important processes ranging from wound healing and tissue development~\cite{Weijer1} to immunological response and diseases such as cancer~\cite{Poincloux}. The best characterized case is that of a crawling cell on a planar 2D substrate or wall. Here motility is generally attributed to cytoskeletal actin filaments which polymerize at one end ($+$) and depolymerize at the other ($-$) in a process called treadmilling. So long as the system has nonzero polarity ${\bf P} = \langle {\bf p}\rangle$ (where ${\bf p}$  is a unit tangent oriented from $-$ to $+$ and angle brackets denote an average over filaments), treadmilling leads to macroscopic motion. This exploits a Brownian ratchet mechanism in which forward fluctuations of the cell perimeter are locked in by polmerization~\cite{Ratchet}. However this mechanism requires a relatively solid anchor-point from which to propel the cell. For cells crawling on a 2D surface, this is provided by focal adhesions and other integrin rich structures~\cite{Theriot-Kondev,Theriot-Mogilner}.

\begin{figure}[h]
\begin{centering}
\includegraphics[width=1.0\columnwidth]{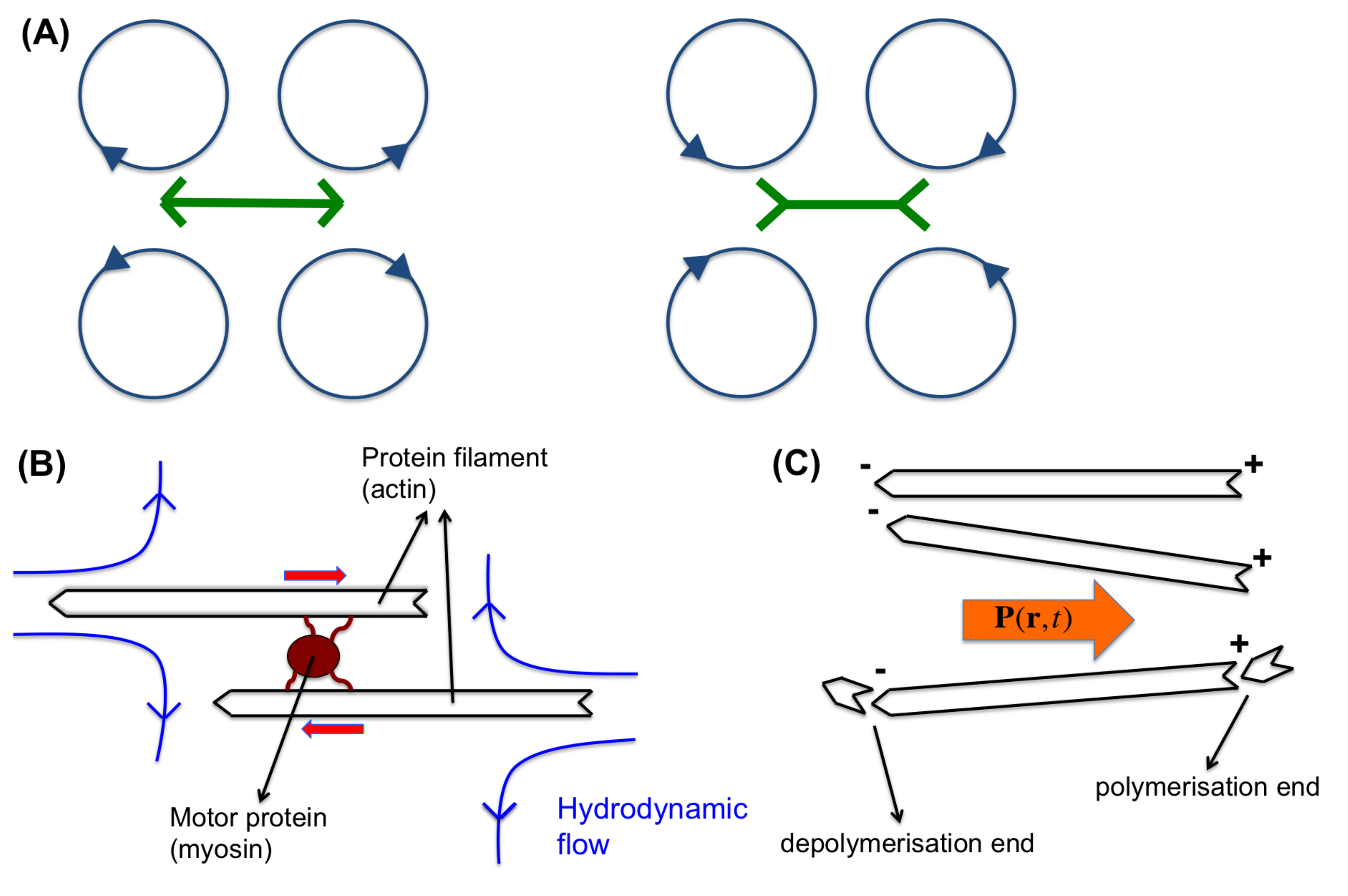}
\end{centering}
\caption{\textbf{(A)} A minimal model for an active element or force dipole.
The figure shows the quadrupolar nature of the hydrodynamic flow around an extensile element or pusher (left)
and a contractile element or puller (right).
\textbf{(B)} 
Contractile stress is created when a motor protein (myosin) pulls protein
filaments (actin) together in the cytoskeleton. 
\textbf{(C)}
Polymerization of the actin filaments gives rise to an effective ``self advection" velocity
in the direction of the polarisation vector (orange arrow). \label{fig:mechanism}}
\end{figure}

In vivo, cells often move in a three-dimensional environment such as an extracellular matrix and/or a tissue of cells~\cite{Poincloux,Evan-Ram,Hawkins}. 
Especially in quasi-spherical motile cells,
both integrin-rich structures and mechanical anchoring are less in evidence, 
and the mechanisms of motility in such 3D environments remain unclear~\cite{Friedl}.
A recent study on tumour cells moving inside an elastic gel suggests that an important role in 3D locomotion may be played not by polymerization but by myosin contractility. This can lead to collective internal flows of actin that may ultimately propel the cell forward~\cite{Hawkins}. The contractile effect arises by a motor spanning two fibres causing an inward force pair (Fig.~\ref{fig:mechanism}A)~\cite{Liverpool}.  This creates an active stress, usually modelled as $\sigma_{\alpha\beta}^{active}=\zed cP_{\alpha}P_{\beta}$ with $c$ the concentration of active material, Greek suffices denote Cartesian directions, and $\zed$ is an activity parameter ($\zed>0$ for contractile systems).
This raises an important issue of principle: how exactly does a tensorial active stress result in a vectorial propulsion velocity?

Several studies use minimal models to address the fundamental physics of how activity imparts cell motility. Experimental progress has focused on ``cell extracts''~\cite{Carlier,Bausch}: unregulated bags of cytoskeletal filaments (actin) and molecular motors (myosin), enclosed by a membrane. However, most of this work focuses on 2D crawling via the treadmill-ratchet mechanism described above. (3D systems are harder to study, and selective inhibition of the treadmill dynamics is biochemically difficult~\cite{Gerisch}.) On the modelling side, generic theories have been proposed to make contact with the cell-extract data, again focusing mainly on 2D crawling and the treadmill-ratchet mechanism~\cite{treadmilling,Wolgemuth,Ziebert}.
While the influence of myosin contractility on cell shape during locomotion has been addressed~\cite{Wolgemuth,Ziebert}, in 2D this has not so far been thought sufficient by itself to lead to motility.

Here we provide a detailed computational study of the effects of active stresses in a minimal 2D model of an actomyosin cell extract. To confirm that our proposed motility mechanism remains pertinent in 3D, we additionally perform selective (computationally intensive) simulations in that case.
Our simulation model comprises a droplet of an active fluid or gel~\cite{Kruse,Tjhung}, confined by interfacial tension $\tilde{\sigma}$, 
and surrounded by a Newtonian host fluid.  Our model equations are based on established continuum precepts and outlined in Methods and Materials. Myosin contractility is represented by an active stress as detailed above. This term is invariant under global
polarity inversion as are, with treadmilling absent, the full equations of motion
(for further discussions of their symmetries, refer to Appendix).
We show then that when the activity parameter $\zed$ exceeds a given threshold, an initially circular or spherical droplet spontaneously breaks that inversion symmetry, leading to an elastic splay of the polarity field and to motion 
along $\pm{\bf P}$. This spontaneous symmetry breaking (SSB) manifests itself as a supercritical Hopf bifurcation, which can alternatively be viewed as  a continuous nonequilibrium phase transition. (The threshold value depend on both 
$\tilde{\sigma}$, 
and an effective elastic constant $\kappa$ penalising distortions of the ordered polar state.) 
In 2D, the resulting motile droplets have crescent-like 
shapes similar to some crawling cells~\cite{Theriot}, 
whereas in 3D our model predicts both spherical and concave shapes. Representing wall friction by a depth-averaged drag term in 2D, we find this slows down the SSB-mediated motion, but does not stop it
altogether unless a critical drag is exceeded.

We then introduce the treadmilling effect, which we represent by a self-advection parameter $w$. 
This breaks the 
global $\mathbf{P}\leftrightarrow-\mathbf{P}$ symmetry, but it does so directly, not spontaneously. (Note that in our model polarity is present even without treadmilling: see Appendix A for a discussion of the relevant physics.) 
This $w$ is the speed at which, relative to the local suspending fluid (of velocity $\mathbf{v}(\mathbf{r},t)$), each filament is self-propelled along its own tangent. Even though asymmetric polymerization does not lead directly to mass transport, for actomyosin the resulting mass flux $wc\mathbf{P}$ should capture, in a highly simplified manner, the preferential growth of filaments by addition of monomers at one end and loss at the other; see Fig.~\ref{fig:mechanism}C. 
This simplified description is possible because we exclude the bath of monomers from the local mass density $c(\mathbf{r},t)$ of active material. (We assume that on average, the monomers nonetheless keep up with the moving gel.)
In general, self-advection leads to density gradients, which
in turn cause hydrodynamic flows $\mathbf{v}(\mathbf{r},t)$ in the direction opposite
to $\mathbf{P}(\mathbf{r},t)$. This backflow severely limits the effectiveness of self-advection in creating motility, almost cancelling it for a droplet in free space as we discuss below. Our 2D study of the effect of wall friction shows however that high enough friction, by reducing $\mathbf{v}$ towards zero, restores treadmilling-induced motility with a speed that approaches $w\mathbf{P}$. 

Although our main focus is on actomyosin cell extracts, our framework provides a broader generic approach to droplet motility. We can thus investigate what happens when we reverse the sign of the active stresses to consider the extensile case, $\zeta = -\zed >0$~\cite{Baskaran,SoftMatter}. The primary experimental relevance of this case is to suspensions of bacteria, which push fluid out along their major axes and draw it in around the equator; see Fig.~\ref{fig:mechanism}A. 
In this context, $w$ is the bacterial swim speed; although this is nonzero for motile species, one may create virtually non-motile mutants (called ``shakers") which still create extensile active stresses, e.g. by excessively increasing the tumbling rate~\cite{tumblers}. A droplet of such organisms can be created either by inducing an attraction between them (as our model effectively assumes)~\cite{jana} or perhaps by confining them in an emulsion droplet. To attain nonzero $\mathbf{P}$ one further requires these particles to have net polar order (as opposed to a nematic state, for which there is orientational order, but equal numbers of particles with tangent $\pm\mathbf{p}$ locally
(see Appendix)). Setting aside the possible difficulties in meeting all those requirements experimentally, we predict that such ``shaker" bacterial droplets could again break the symmetry and start moving spontaneously. Intriguingly, the predicted trajectories  in this extensile case are more complicated than those of the contractile model. This is because, in extensile droplets, the SSB-mediated velocity forcing is in a direction perpendicular,
rather than parallel, to the polarization vector $\mathbf{P}$. 
Restoring nonzero $w$ to describe the case of motile bacteria, the composition of these two motions leads to circular or spiralling trajectories of the droplet as a whole.

\section{Results}

We first present results for the 2D contractile case, then briefly describe our findings in 3D, and finally give some further 2D results on extensile droplets. We initialized our simulation runs with a circular (or spherical) droplet within which the concentration of active material is taken to be a constant ($c=c_0$), with $c=0$ outside. The polarization field $\mathbf{P}(\mathbf{r},t)$ within the droplet is initially uniform along the horizontal ($\hat{x}$) axis and varies with concentration as  
magnitude $\sqrt{\frac{(c-c_{cr})}{c_{cr}}}$ with $c_{cr}$ a threshold value for polarization onset. {We choose $c_0>c_{cr}>0$ so that $|\mathbf{P}|$ is nonzero inside the droplet and zero (isotropic) outside, so that we assume the actomyosin network inside our active droplets is initially polarised.}
The system is then evolved via the equations of motion as specified in Materials and Methods {(discussed further in Appendix A)}, with chosen values of the activity parameter $\zed$ and the self-advection (treadmilling) parameter $w$. 
On a relatively short timescale both the internal concentration and the polarisation field relax towards the equilibrium values $c_{eq}$ and $\mathbf{P}_{eq}$
(while still $c=0$ and $\mathbf{P}=\mathbf{0}$ externally)
with some interfacial tension $\tilde{\sigma}$, set by minimization of our chosen free energy.
(Our choice creates no anchoring of $\mathbf{P}$ at the surface so, in the absence of symmetry breaking, the polarization remains uniform; see Fig.~\ref{fig:splay}A left.
{The effect of a soft anchoring is discussed in the Appendix A.}) 
Having made one such relaxed droplet, the dependence of its behavior on $\zed$ and/or $w$ was systematically explored by incrementing those quantities and waiting for steady state, before incrementing again. Because we are primarily interested in trends and symmetry breaking phenomena, rather than quantitative predictions of where these will occur for specific materials, we report all results below in the natural units for lattice Boltzmann simulations (LBU); the connection between these and physical units is discussed in Appendix D.
 
\subsection{Contractile stress can create motility via SSB}

\begin{figure}[h!]
\begin{centering}
\includegraphics[width=1.0\columnwidth]{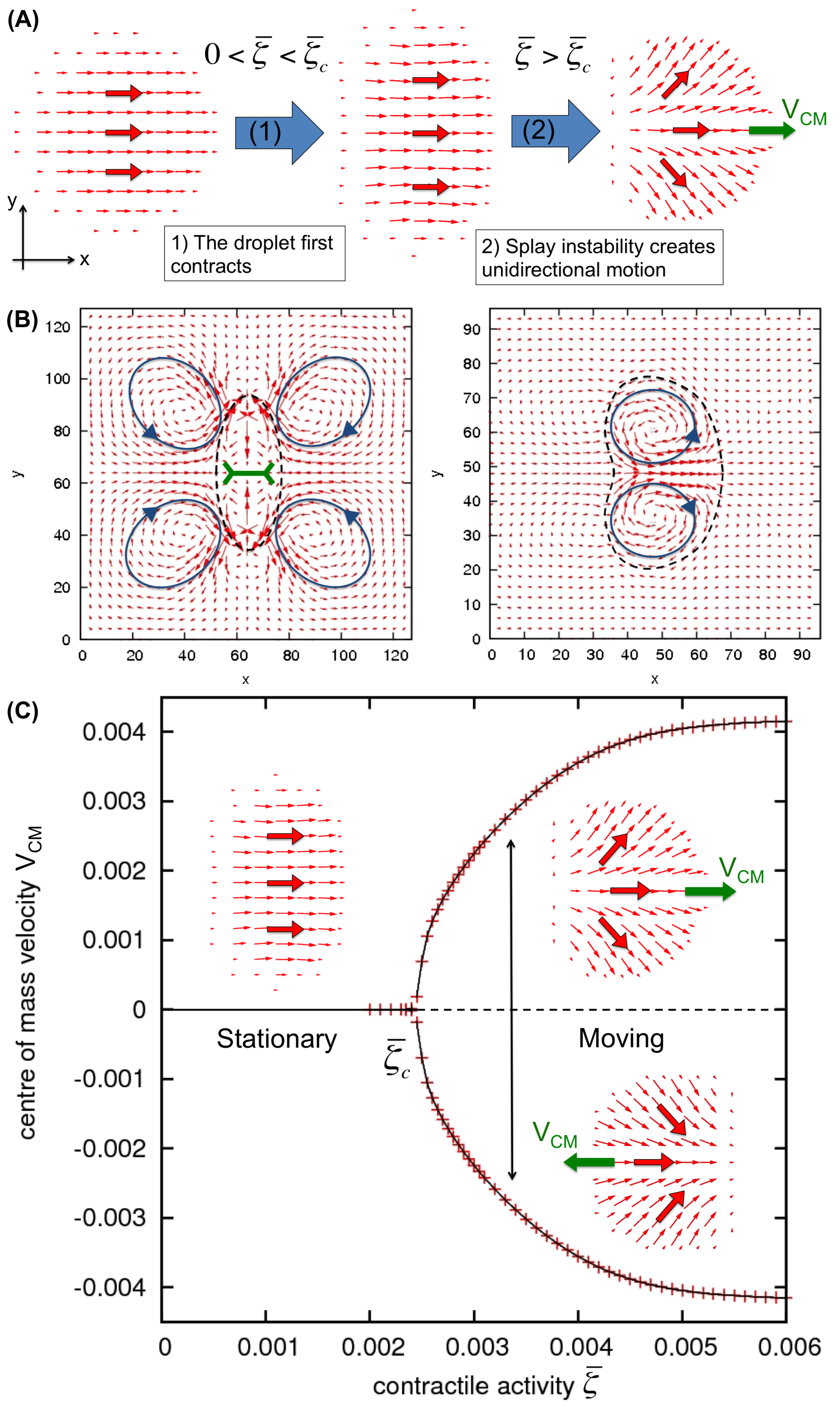}
\end{centering}
\caption{\textbf{(A)} 
Steady state configurations of a contractile active droplet without self-advection. 
The (red) arrows show the polarisation field $\mathbf{P}(\mathbf{r},t)$. Upon increasing the contractile activity $\zed$, the droplet elongates perpendicular to $\mathbf{P}$ and then becomes unstable with respect to splay deformation at critical activity $\zed_{c}$. 
When it splays, the droplet also spontaneously moves in the direction of the 
green arrow. The time evolution of the system 
is shown in Supplementary Movie 1.
\textbf{(B)}
Left plot shows the velocity field of the droplet at $\zed<\zed_c$ which is quadrupolar, like that around a contractile element 
(Fig. \ref{fig:mechanism}A right). 
Right plot shows the velocity field of the splayed and moving active droplet which consists of two opposing vortices. 
The boundary of the droplet itself is given by the dashed line. 
\textbf{(C)}
Bifurcation diagram showing spontaneous symmetry breaking from a uniform
and stationary state to a splayed and moving state as the activity parameter $\zed$ is increased.
\label{fig:splay} }
\end{figure}

Although contractile motor stress and actin treadmilling are generally both present in motile cells, it is illuminating to 
study these two mechanisms separately. We first consider a droplet with no external drag term (no wall friction) and no treadmilling term ($w=0$), and vary the activity parameter $\zed>0$.
For low activity $\zed$, below some critical value $\zed_{c}$,
the droplet polarisation field $\mathbf{P}$ remains aligned uniformly along its initial direction  $\hat{x}$
and the droplet remains stationary. However it becomes slightly
elongated in the direction perpendicular to the polarisation vector
$\mathbf{P}$ as a result of the competition between the contractile
stress and the interfacial tension (see Fig.~\ref{fig:splay}A
middle). In this regime, the contractile stress set up a quadrupolar fluid
flow around the droplet (Fig.~\ref{fig:splay}B left), so that the whole droplet behaves as a large contractile element  (compare Fig.~\ref{fig:mechanism}A right). However it does not translate: there is no motility, and the droplet is a ``shaker" rather than a ``mover"~\cite{Ramaswamy}.

As we increase $\zed$ beyond $\zed_{c}$, the uniform polarization field $\mathbf{P}$ becomes unstable with respect to a splay deformation. This happens because the contractile stress is large enough to overcome the resistance to deformation mediated by the elastic constant $\kappa$. The splay creates a state in which neighboring vectors $\mathbf{P}$ either fan outwards ($\nabla\cdot\mathbf{P}>0$) or inwards ($\nabla\cdot\mathbf{P}<0$). The first is shown in Fig.~\ref{fig:splay}A right; the second is found by first taking its mirror image and then reversing $\mathbf{P}$. This choice is made at random, spontaneously breaking the
{global}
polarity inversion symmetry. As soon as this happens, the droplet starts to move along the direction set by  $(\nabla\cdot\mathbf{P})\mathbf{P}=\pm\mathbf{P}$. This motion is attributable to the formation of a pair of flow vortices inside the droplet (Fig.~\ref{fig:splay}B right). 
{
Such spontaneous propulsion is somewhat reminiscent of the self-electrophoretic motion of a vesicle with active membrane pumps in an ionic solution~\cite{Mitchell,Lammert}.} 

We note that in bulk active contractile
fluids, the state of uniform $\mathbf{P}$ is also generically unstable to splay fluctuations in 1D~\cite{Baskaran,Simha,Giomi-Marchetti}, which then lead to the
onset of spontaneous flow. {At one level, the SSB-induced motility transition described here can be viewed as a manifestation of that bulk instability, albeit with two variations.} First, spontaneous flows are present on both sides of our transition: as discussed above there is a quadrupolar flow field already in the non-moving state. Second, in bulk the critical activity level is nonzero only in finite systems, for which the transition is discontinuous, unlike ours (see below and Fig.~\ref{fig:splay}C), and the resulting velocity field 
much more complicated~\cite{Tjhung,SoftMatter}.

To illustrate our symmetry
breaking motility mechanism more clearly, we plot the magnitude $V_{CM}$ of the centre of mass velocity of
the droplet $\mathbf{V}_{CM}$ as a function of activity $\zed$ in Fig.~\ref{fig:splay}C. This bifurcation diagram shows a continuous non-equilibrium transition
from a stationary and uniform state to a moving and splayed state. Moreover, to within numerical accuracy the observations are consistent with a supercritical
Hopf bifurcation, for which $V_{CM}\sim(\zed-\zed_{c})^{0.5}$. This mean-field like exponent is perhaps unsurprising as there is no noise in our simulations. Accordingly it might change in the presence of activity-generated noise~\cite{harvardSGR}, depending on whether the bifurcation remains low-dimensional or acquires a many-body critical  character. 

\subsection{SSB-induced motility is diminished by friction}

\begin{figure}[h!]
\begin{centering}
\includegraphics[width=0.9\columnwidth]{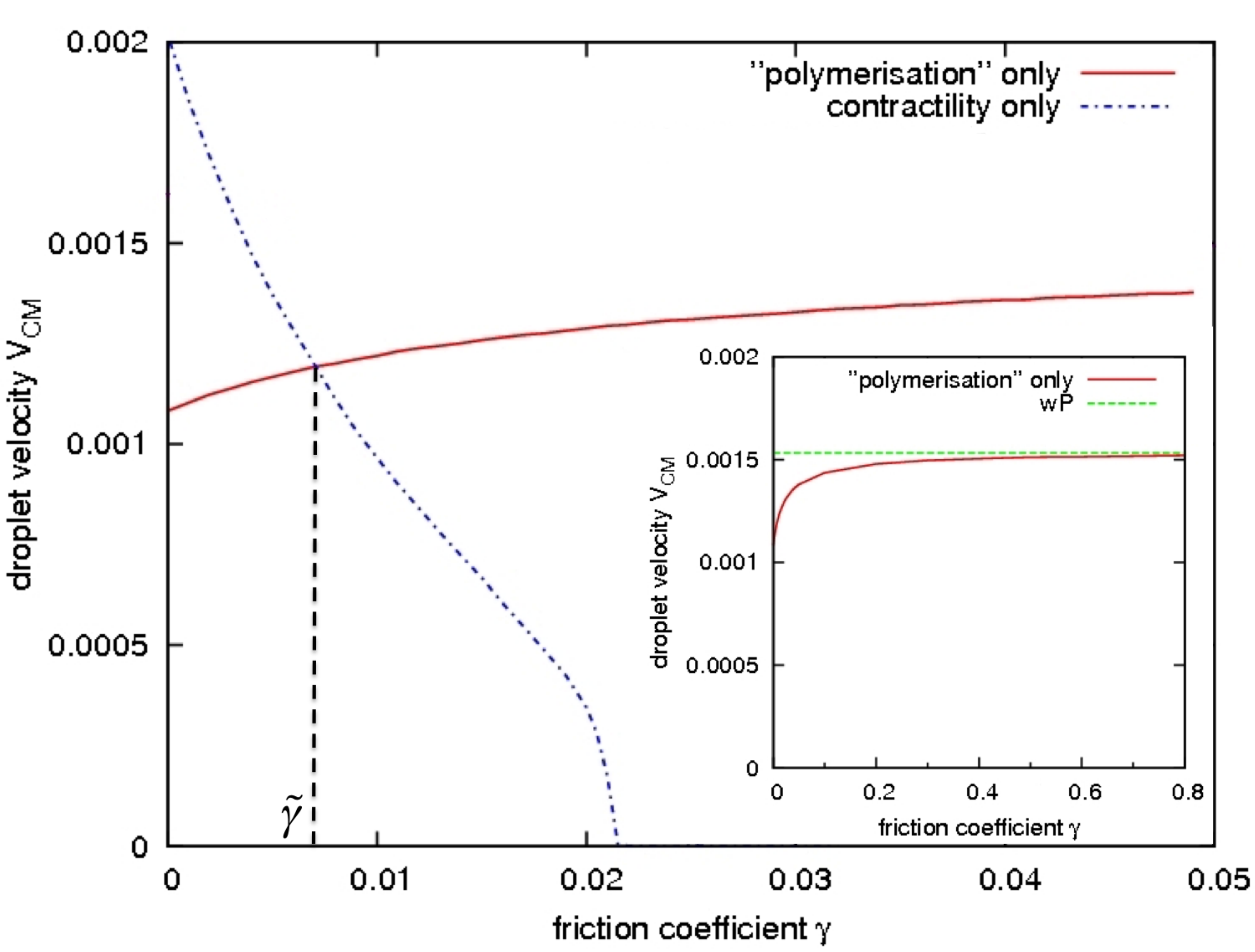}
\end{centering}
\caption{
Representative plots of droplet velocity $V_{CM}$
against frictional coefficient $\gamma$ for motile droplets driven by:
contractile stress only (solid red line) and polymerization/self-advection only (dashed blue line).
The inset shows the polymerisation-only driven motility in the limit of large friction in which the droplet velocity
approaches ``polymerisation'' speed $wP$.
\label{fig:friction}}
\end{figure}

The results above are for a droplet in 2D surrounded by Newtonian fluid. To better describe experiments involving cell-crawling on a substrate, we now consider an additional frictional force between the solid wall
and the cell.  To do this within our 2D continuum model, we introduce an additional force density $\mathbf{f}_{friction}=-\gamma\mathbf{v}$ to the momentum balance equation (see Materials and Methods and Appendix). Here $\gamma$ is an effective coefficient of friction which depend on whether we have non-slip or partial-slip boundary conditions on the substrate and also on the thickness
of the cell. It may represent conventional friction and/or a coarse-grained model of focal adhesions and other localized mechanical contacts. The presence of this friction will significantly quench
the hydrodynamic flow $\mathbf{v}$. Since our SSB-induced motility requires a hydrodynamic vortex flow inside the
droplet, the frictional force can greatly reduce the droplet's velocity, bringing it to rest beyond a critical level $\gamma_c$, at which point the symmetry is restored; the value of $\gamma_c$ depends on activity and other model parameters.
This is seen in a plot of  $V_{CM}$ against $\gamma$ in Fig.~\ref{fig:friction}. 

\subsection{Self-advective motility is enhanced by friction}

We now consider the case when there is only treadmilling (modeled as self-advection $w>0$)
but no contractile stress ($\zed=0$). As discussed previously, the droplet will move along the polarization vector $\mathbf{P}$ with speed
$V_{CM}<w$. Since there is no spontaneous symmetry breaking
involved, translational motion occurs for any nonzero $w$, in contrast to the threshold behavior seen for contractile SSB-motility.
The most interesting aspect is the role of the friction parameter $\gamma$. 
Again we plot the velocity of the droplet $V_{CM}$ as a function
of $\gamma$ in Fig.~\ref{fig:friction}. In contrast to the previous case, motility is significantly enhanced by the presence of friction. Indeed, in in the limit $\gamma\rightarrow\infty$, we have $\mathbf{v} = \mathbf{0}$ and $\mathbf{V}_{CM} \to w\mathbf{P}$ 
(see Fig.~\ref{fig:friction} inset).

The intersection of the two plots of droplet velocity versus friction (found respectively by switching off activity or self-advection) defines a characteristic friction scale 
$\widetilde{\gamma}$.  For $\gamma<\widetilde{\gamma}$
contractile stresses dominate cell motility, while for $\gamma>\widetilde{\gamma}$, self-advection is dominant. 

In most experiments on 2D crawling of cells/cell extracts
\cite{Theriot-Mogilner}, 
the involvement of focal adhesions suggests that the high friction (treadmilling dominated) limit generically prevails. On the other hand,
some recent experiments~\cite{Yam} directly identify spontaneous symmetry breaking of the actomyosin network as the initiator of polarized cell motility in keratocytes. Our work emphasizes that {\em spontaneous} breaking of 
{global}
polarity inversion symmetry arises from contractile motor activity, not from treadmilling. It is therefore  arguable that the role of motor activity in 2D motility has so far been underestimated.
We note however that an equivalent discrete symmetry breaking would create motility if a pure treadmilling state of zero $P_\alpha=\langle p_\alpha\rangle$, but finite nematic order ($\langle p_\alpha p_\beta\rangle - \delta_{\alpha\beta}/3 \neq 0$),
spontaneously acquires polarity 
{locally (see Appendix)}.

\begin{figure}[h!]
\begin{centering}
\includegraphics[width=1.0\columnwidth]{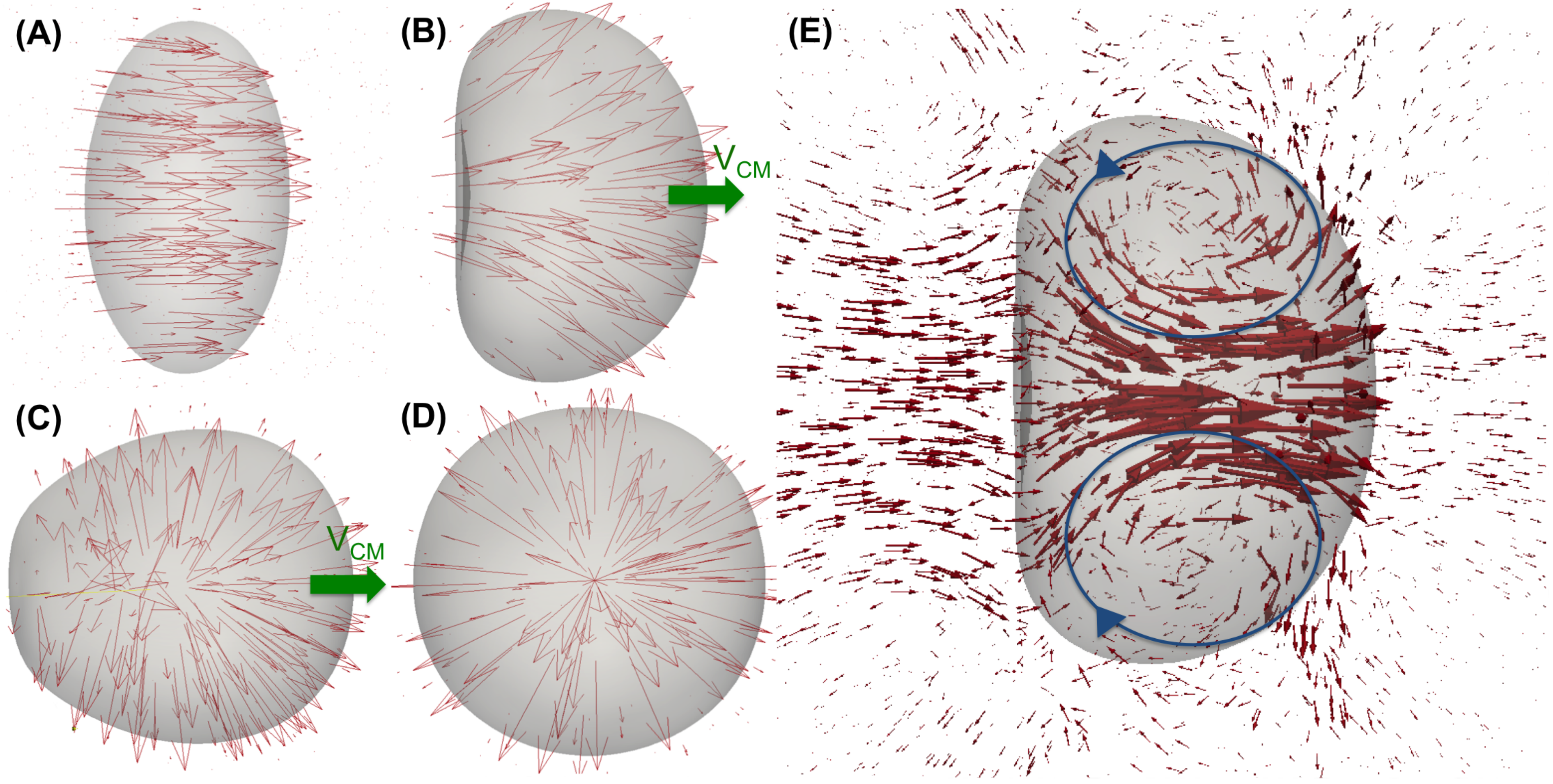}
\end{centering}
\caption{Steady state conformations in 3D contractile droplets without self-advection
on increasing activity $\zed$ from \textbf{(A)} to \textbf{(D)}.
B and C are motile as indicated while
A and D are stationary.
\textbf{(E)} shows the toroidal fluid flow inside the motile droplet of steady state B. The time evolution of the droplets in \textbf{(B)} and \textbf{(D)}
are shown in Supplementary Movies 2 and 3 respectively.
\label{fig:3D}}
\end{figure}

\subsection{3D droplets show a window of SSB-induced motility}

In the context of experiments on 3D tumour cells, it has been argued that motility is driven primarily by contractile stress~\cite{Poincloux,Hawkins}, suggesting that the low-friction limit of our model prevails here. This accords with the much diminished part played by focal adhesions in 3D
\cite{Friedl}. 
For the 3D case we therefore neglect the friction term, and run selected simulations to confirm that the SSB route to motility remains operative. 

Fig.~\ref{fig:3D} shows steady state polarization fields inside a 3D contractile droplet 
with increasing values of the activity magnitude from A to D. 
As expected, the first steady state encountered is a symmetric but
deformed immotile droplet (Fig.~\ref{fig:3D}A). The active stress contracts the droplet along $\pm\mathbf{P}$ resulting in a lenticular shape.
As we increase $\zed$ beyond a critical $\zed_c$, splay instability
spontaneously breaks symmetry and causes the droplet to move along $(\nabla.\mathbf{P})\mathbf{P}$ just as in the 2D case. The droplet shape is concave and (as in the lenticular case) both it and the flow field resemble that created by rotating the 2D droplet about the $\mathbf{P}$ axis (Fig.~\ref{fig:3D}B). 
The resulting hydrodynamic flow (Fig.~\ref{fig:3D}E) therefore corresponds to a toroidal vortex ring.
It would be interesting to see how these predictions
compare with intracellular actin and fluid flow maps that might in future be measured for cells moving in 3D environments, 
for instance those studied in Ref.~\cite{Poincloux}. 

Interestingly, as $\zed$ is increased further, the droplet becomes increasingly spherical (Fig.~\ref{fig:3D}C), and finally symmetry is restored, creating an immotile spherical droplet with a `hedgehog' defect (of topological charge 1 as dictated by the polar ordering~\cite{Chaikin-Lubensky}) 
at the centre (Fig.~\ref{fig:3D}D).

\subsection{Extensile SSB creates transverse or circular motility}

Our final results are for 2D extensile droplets (Fig.~\ref{fig:mechanism}A). Continuum descriptions of uniform extensile active fluids are widely used to describe dense bacterial suspensions~\cite{Baskaran,Ramaswamy,Suzanne}. These results may therefore be relevant to bacterial droplets formed by aggregation in the presence of attractive forces~\cite{jana}, or possibly by confinement of bacteria within a droplet emulsion. Again, for fixed thermodynamic parameters controlling the elasticity and interfacial tension of the confined material, one can vary the
activity parameter $\zeta = -\zed$ (now positive) and the self-advection
parameter $w$ which corresponds to the swimming speed of individual bacteria and is nonzero for ``movers" but zero for ``shakers" ~\cite{Baskaran,Simha}.

\begin{figure}[h!]
\begin{centering}
\includegraphics[width=1.0\columnwidth]{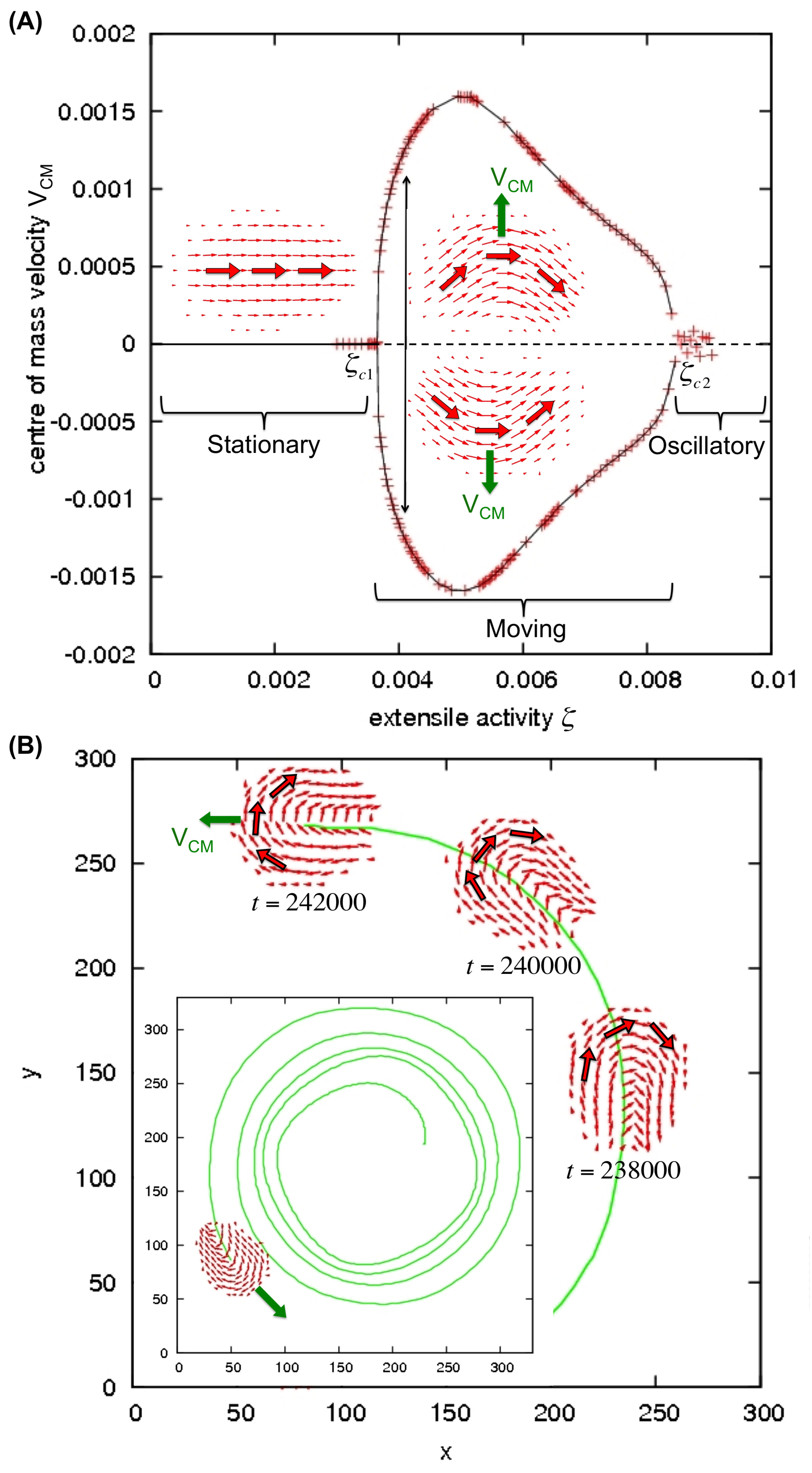}
\end{centering}
\caption{ 
\textbf{(A)} 
Plot of centre of mass velocity against activity ($\zeta=-\zed>0$) for extensile droplet without self-advection.
It shows continuous transitions from stationary to motile and then from motile to oscillatory at critical activity
$\zeta _{c1}$ and $\zeta _{c2}$ respectively.
Also shown are the steady state polarisation field $\mathbf{P}$ for the stationary and motile case. The movie of the time evolution of the system is shown
in Supplementary Movie 4.
\textbf{(B)} 
The presence of both extensile stress and self-advection leads to an outward spiral trajectory (solid green lines).
Also shown are the snapshots of the polarisation field at different timesteps (red arrows).
\label{fig:extensile}}
\end{figure}

Fig.~\ref{fig:extensile}A shows steady state configurations of a 2D purely extensile ($w=0$) droplet
at different ranges of activity.
For $\zeta<\zeta _{c1}$ the droplet remains stationary but again elongates symmetrically, this time {\em along} the
direction of the polarization field $\pm{\mathbf P}$. The extensile stress creates the quadrupolar flow field expected of a large, extensile shaker.
For $\zeta$ beyond the critical value $\zeta _{c1}$,
the droplet again becomes unstable,  
but now with respect to bend deformation as opposed to splay. 
This gives rise to a horizontal vortex pair inside the droplet as opposed to the vertical one in the contractile case (Fig.~\ref{fig:splay}B right). This flow field causes the droplet to move in a direction set by the sense of bending as ${\mathbf P}\times\left(\nabla\times {\mathbf P}\right)$ 
(which can be upwards or downwards according to Fig.~\ref{fig:extensile}A).
As can be seen from figure Fig.~\ref{fig:extensile}A this stationary-to-motile transition can still be characterised as a continuous SSB transition, however, the droplet speed $V_{CM}$ attains a maximum at intermediate $\zeta$ before falling to zero again.
Beyond a second critical activity threshold ($\zeta _{c2}$ in Fig.~\ref{fig:extensile}A), the polarisation pattern oscillates continuously while remaining symmetric at all times and consequently, the droplet again becomes non-motile. 

Turning finally to the case of nonzero self-advection ($w>0$) we find that this, combined with the extensile motility, can give rise to an intriguing spiralling 
motion (see Fig.~\ref{fig:extensile}B).
This arises because the SSB-induced motility is at right angles to the polarization (Fig.~\ref{fig:extensile}B right).
On the other hand, self-advection everywhere transports material along $\mathbf{P}(\mathbf{r},t)$ locally; since the polarization field is spontaneously curved, this by itself would give a circular droplet orbit. Compounding these two motions typically leads to outward spiral trajectories 
as shown in Fig.~\ref{fig:extensile}B. This outcome contrasts with the contractile case where both the SSB-induced and the self-advective motion (the latter averaged over the droplet configuration) point either together or oppositely along the $\pm\mathbf{P}$ direction and only straight line motion can result.

\section{Discussion and Conclusions}

Our simulations of contractile droplets can be viewed as a simple {\it in silico} analogue of the {\em in vitro} cell-extract studies that have been used to dissect the biophysical ingredients of motility.  Our simulation work powerfully complements these studies, by allowing us to isolate the role of contractile (motor) activity in cell locomotion: this is very difficult in the laboratory, where current strategies for inhibiting polymerization dynamics (treadmilling) can severely impair other key subcellular processes~\cite{Gerisch}. 

Our approach likewise complements those of previous theories~\cite{Wolgemuth,Ziebert,Doubrovinski} 
which have concentrated on treadmilling as the main driver of motility. We have shown in 2D that contractile 
stresses alone can not only shape the rear of a crawling 
cell (Fig.~\ref{fig:splay}), but also create motility itself, provided that the motor forces are large enough to create an asymmetric circulatory flow as in Fig. \ref{fig:splay}B. 
Interestingly, keratocyte cells crawling on glass seemingly do
exploit myosin activity to set up an intracellular actin flow in the rear of a cell which ultimately polarises it and makes it motile~\cite{Yam}. 
Nonetheless, our study of the the effects of a frictional term, which promotes the motility created by treadmilling but inhibits that caused by contractile stress, lends support to the view that cell crawling on a wall is usually dominated by the treadmilling.

In contrast, in 3D cell motility, recent work suggests that treadmilling plays at most a minor role~\cite{Poincloux,Hawkins}. In the 3D case, therefore, our work describes a simple and compelling mechanism for how spontaneous translational motion can {in principle} arise solely by the action of a contractile stress. This requires spontaneous symmetry breaking, mediated in our case by splay deformation in response to that stress.
Our 2D and 3D simulations go beyond the 2D theory of Ref.~\cite{Hawkins} by addressing the dynamics of the polarization field.
We do however make some important simplifications: our droplets are confined only by interfacial tension not by an elastic membrane; we treat treadmilling as a simple self-advection; and we do not address any direct transition between nematic and polar order, despite assuming polarity inversion symmetry at thermodynamic level. 
Improving the model in these respects will require a more detailed microscopic derivation which we shall leave to future work.
To test whether our model indeed captures the biophysics of 3D cell motility, it would be exciting to visualize experimentally the detailed cytoskeletal organization and flow fields, e.g. for cells moving through matrigel~\cite{Poincloux,Hawkins}.  

{Our generic framework is not
limited to contractile actomyosin networks. Indeed we have discussed the case of extensile droplets, possibly relevant to aggregates or emulsions of active but immotile bacteria; here translational motility arises by spontaneous symmetry breaking only at intermediate activity, and is mediated by bend rather than splay deformation. The addition of self-advection along the bent polarization field then leads in addition to rotational motion. We note that rotational and translational motility of small bacterial aggregates was recently observed, but attributed to a somewhat different mechanism where symmetry is broken by frozen-in statistical fluctuations rather than SSB~\cite{jana}.
}

Finally, our hydrodynamic equations of motion, or close variants of these, might in some cases be applicable to concentrated eukaryotic cell masses such as tissue~\cite{Ranft}.
Within a tissue each cell exerts forces on its neighbors which at the lowest order continuum level creates a certain density of force dipoles~\cite{Ranft}; the
velocity field $\mathbf{v}$ then describes the slow migration of cells  inside the tissue. 
It is intriguing to note that the large-scale tissue flow in animal cells during gastrulation may break the symmetry to form vortices, similarly to our active droplets. This is the case of the so-called ``polonaise movements'' which are observed in the developing chick embryo~\cite{Weijer1}, and which are important to form the correct supercellular structure. In this context a ``polarisation'' field is sometimes used to describe the orientation of individual cells~\cite{Vasiev}.
The relation, if any, between the onset of this vortex flow and that seen in our droplets remains to be explored.

\section{Materials and Methods}

We briefly outline here the hydrodynamic model used in this work (more details are in Appendix). 
We consider a fluid, comprising a mixture of active material and solvent, with 
constant total mass density $\rho$. 
The hydrodynamic variables whose
dynamics we monitor are: (i) the concentration of active material
$c(\mathbf{r},t)$, (ii) the fluid velocity $\mathbf{v}(\mathbf{r},t)$ (with incompressibility requiring $\nabla.\mathbf{v} = 0$), and 
(iii) the polarization field $\mathbf{P}(\mathbf{r},t) = \langle \mathbf{p}\rangle$ as defined previously.  


Although an active droplet is a nonequilibrium system, we introduce the 
following free energy functional to describe its equilibrium physics in the passive limit of zero activity: 
\begin{eqnarray}\label{fe}
F[c,\mathbf{P}] &=& \int d^{3}r\,\{ V(c)+\frac{k}{2}\left|\nabla c\right|^{2}-\frac{\alpha}{2}  \frac{(c-c_{cr})}{c_{cr}}  \left|\mathbf{P}\right|^{2}  \\\nonumber
		&+& \frac{\alpha}{4}\left|\mathbf{P}\right|^{4}+\frac{\kappa}{2}(\nabla\mathbf{P})^{2}\} 
\end{eqnarray} 
Here $\alpha>0$ is a phenomenological free energy amplitude,
$k$ determines the droplet interfacial
tension, and $\kappa$ is an effective elastic constant. This choice of $F[c,\mathbf{P}]$ leads to a 
continuous isotropic-to-polar transition at $c=c_{cr}$.
To confine the active material into a droplet, we choose:
$V(c)=\frac{a}{4c_{cr}^4}c^{2}(c-c_{0})^{2}$
and set $c_0>c_{cr}$. This creates two free energy minima corresponding to a phase of pure passive solvent (external to the droplet, $c=0$
and $\mathbf{P}=\mathbf{0}$)
and a polar active phase (inside the droplet, $c=c_{eq}>c_{cr}$
and $\mathbf{P}=\mathbf{P}_{eq}$). 


Treating the active material as locally conserved, the time evolution
of the concentration field $c(\mathbf{r},t)$ can then be written
as a convective-diffusion equation:
\begin{equation}\label{ceq}
\frac{\partial c}{\partial t}+\nabla\cdot\left(c(\mathbf{v}+w\mathbf{P})-M\nabla\frac{\delta F}{\delta c}\right)=0
\end{equation}
where $M$ is a thermodynamic mobility parameter and $w$ is the self-advection parameter. 

The dynamics of the polarisation field $\mathbf{P}(\mathbf{r},t)$
follows an ``active nematic'' evolution~\cite{SoftMatter}, 
given by
\begin{equation}\label{Peq}
\frac{\partial\mathbf{P}}{\partial t}+\left((\mathbf{v}+w\mathbf{P})\cdot\nabla\right)\mathbf{P}=-\underline{\underline{\Omega}}\cdot\mathbf{P}+\xi\underline{\underline{v}}\cdot\mathbf{P}-\frac{1}{\Gamma}\frac{\delta F}{\delta\mathbf{P}}
\end{equation}
where $\underline{\underline{v}}$ and $\underline{\underline{\Omega}}$
are the symmetric and anti-symmetric parts of the velocity gradient
tensor $\nabla\mathbf{v}$. $\Gamma$ is the rotational viscosity and $\xi$ is related to the geometry
of the active particles~\cite{Kruse}.

{Force balance is ensured through the 
Navier-Stokes equation,
$\rho\left({\partial}/{\partial t}+\mathbf{v}\cdot\nabla\right)\mathbf{v}  =  -\nabla P+\nabla\cdot\underline{\underline{\sigma}}^{total}-\gamma\mathbf{v},
$
}
where $P$ is the isotropic pressure, $-\gamma\mathbf{v}$ is the friction force per unit volume and 
$\underline{\underline{\sigma}}^{total}$
is the total stress in the fluid which includes viscous, elastic/Ericksen, interfacial and ``active'' stresses {(see Appendix for details and for a
discussion of the effects of additional terms, allowed by symmetry, in the 
equations of motion).}
The active stress is  
$\sigma_{\alpha\beta}^{active}=\zed cP_{\alpha}P_{\beta}$ \cite{Ramaswamy}
where $\zed >0$ for contractile activity and $\zeta = -\zed>0$ for extensile.

To solve these equations in 2D and 3D, we performed hybrid lattice
Boltzmann simulations, as done previously for other active flows~\cite{Tjhung,SoftMatter}. 

\begin{acknowledgments}
We thank R. Voituriez for very useful discussions. ET thanks SUPA for a Prize Studentship and MEC holds a Royal Society Research Professorship. 
\end{acknowledgments}

\appendix

\section{Hydrodynamic description}

\subsection{Choice of order parameters} We model actin filaments in the cytoskeleton, or bacteria in suspension, as rod-shaped active polar particles.
In the hydrodynamic (continuum) limit, the dynamics of these dense suspensions of active
polar particles can be described by a few continuum variables.
 
Our chosen hydrodynamic variables are: the concentration of the active particles
$c(\mathbf{r},t)$; the average velocity field $\mathbf{v}(\mathbf{r},t)$
of both the active particles and the solvent; and finally the polarisation
field $\mathbf{P}(\mathbf{r},t)$ 
which is defined as a mesoscopic average orientation of the
polar particles: $\mathbf{P}=\left<\mathbf{p}\right>$ (with the average taken over molecular orientations $\mathbf{p}$; see  main text).
The total mass density is assumed to be constant throughout
(so the fluid is incompressible). 

Importantly, in this paper we assume $\mathbf{P}$ to be nonzero, even in the absence of self-advection ($w=0$). We believe this to be the correct description for actomyosin gels in a cytoskeletal context: that is, in any mesoscopic region containing $N$ aligned actin fibres, we assume an excess of one orientation over the other that scales as $N$, not as $N^{1/2}$ (as would arise for random orientations) {\cite{PGG}}. Moreover, relaxation of $\mathbf{P}$ in the absence of activity requires slow reversal of filament directions; therefore the local polarization will not suddenly disappear if all activity is abruptly switched off. The resulting presence of a quasi-static polarization is important in principle, because it restricts the nature of the allowed orientational defects in the system to those of integer topological charge \cite{Chaikin-Lubensky}. 

An alternative theory would suppose that without self-advection the system is generically in a nematic state, which has $\mathbf{P} = 0$ but a nonzero tensor order parameter 
$\underline{\underline{Q}}=\left< \mathbf{p}\mathbf{p}-\frac{1}{3} \mathbb{I}\right>$. This describes a different situation in which half-integer defects become possible (at least when $w=0$). A brief discussion of the onset of spontaneous motility in a nematic model is given in Appendix B. Meanwhile we emphasize that orientation of actin, as distinct from self-advection of any kind, is what the polarization field $\mathbf{P}$ represents in our model. 

{Note that in some parts of the liquid crystal literature~\cite{Frank,Ericksen,Yeomans}, the term ``nematic'' is used for a system whose mathematical description involves a nonzero vectorial order parameter (say $\mathbf{P}$), but whose governing equations respect the global symmetry 
\begin{equation}
\mathbf{P}(\mathbf{r})\rightarrow-\mathbf{P}(\mathbf{r})\quad{ \forall\; \mathbf{r}}
\label{global}\end{equation}
}
(as will arise for $\beta = w = 0$ in the equations to be developed below). We believe this terminology {{in our specific problem}} is best avoided: it gives the impression that the spontaneous breakdown of this global symmetry is equivalent to a genuine transition from nematic ($\underline{\underline{Q}}\neq 0, \mathbf{P} = 0$) to polar ($\mathbf{P} \neq 0$). But mathematically there is no such equivalence. Indeed, the nematic-to-polar transition is from a state in which half-integer defects are allowed, to one in which they are forbidden. In contrast, spontaneous breakdown of the symmetry Eq.~\ref{global} represents a transition between two states that both forbid half-integer defects. Put differently, the global symmetry in Eq.~\ref{global} is not equivalent to a local symmetry, in which one can freely reverse the orientation $\mathbf{p}$ of all rods within {\em any} arbitrary subvolume. That {\em local} symmetry is the defining feature of the nematic state. {It is fully built into any description based on the tensor order parameter $\underline{\underline{Q}}$, but the local symmetry is hidden if one replaces this with a director field ${\bf n}$ (a unit vector whose orientation is defined by the major principal axis of 
$\underline{\underline{Q}}$). For instance, a fictitious discontinuity (domain wall) between two slabs with $n_x = \pm 1$ (say) correctly has zero energy cost in the tensorial description but acquires a spurious, and indeed infinite, square gradient energy $\sim\kappa\int (\nabla {\bf n})^2 dV$ in the director field approach. The corresponding domain wall would, in contrast, be physically real in a system with polar order.}

\subsection{Hydrodynamic equations} 

 The time-evolution of hydrodynamic
variables in active systems can either be derived analytically~\cite{Baskaran} or phenomenologically~\cite{Ramaswamy} -- we will follow the latter avenue in
this work.

In deriving phenomenologically the hydrodynamic equations, we first introduce the free energy functional:
\begin{eqnarray}\label{feappendix}
F[c,\mathbf{P}] &=& \int d^{3}r\,\{ V(c)+\frac{k}{2}\left|\nabla c\right|^{2}-\frac{\alpha}{2}  \frac{(c-c_{cr})}{c_{cr}}  \left|\mathbf{P}\right|^{2}  \\\nonumber
		&+& \frac{\alpha}{4}\left|\mathbf{P}\right|^{4}+\frac{\kappa}{2}(\nabla\mathbf{P})^{2}+\beta\mathbf{P}\cdot\nabla c\}. 
\end{eqnarray}
In the passive limit the system tends to minimise this free energy.
Note that we have assumed the single elastic
constant ($=\kappa$) approximation. Furthermore, within our theory in the
passive limit there is a second order transition from
isotropic ($\left|\mathbf{P}\right|=0$) to polar ($\left|\mathbf{P}\right|>0$)
phase at critical concentration $c_{cr}$. Other phenomenological
parameters are $\alpha$, which controls the isotropic-to-polar transition,
and $k$, which in conjunction with $V(c)$ determines the interfacial tension.

To enable droplet formation, we choose:
\begin{equation}
V(c)=\frac{a}{4c_{r}^{4}}c^{2}(c-c_{0})^{2}
\end{equation}
and set $c_{0}>c_{cr}$. This creates two free energy minima corresponding
to a phase of pure passive solvent (external to the droplet, $c=0$
and $\mathbf{P}=0$) and a polar active phase (inside the
droplet, $c=c_{eq}>c_{cr}$ and $\mathbf{P}=\mathbf{P}_{eq}$). Then $c_{eq}$ and $\mathbf{P}_{eq}$ are found
by minimizing $F[c,{\mathbf P}]$ in a state of uniform $c$ and $\mathbf{P}$. The
resulting surface tension $\tilde{\sigma}$ will depend on $a$, $k$,
and the elastic constant $\kappa$. In the special case of
$c_{cr}=c_{0}/2$, the expression for $c_{eq}$ and $\mathbf{P}_{eq}$
are:
\begin{eqnarray}
c_{eq} & = & \frac{c_{0}}{2}+\frac{c_{0}}{2}\sqrt{1+\frac{\alpha}{2a}}\\
\mathbf{P}_{eq} & = & \left(1+\frac{\alpha}{2a}\right)^{1/4}\hat{\mathbf{p}}
\end{eqnarray}
where $\mathbf{\hat{p}}$ is a unit vector.

Finally, the last term in Eq.~\ref{feappendix}, $\beta\mathbf{P}\cdot\nabla c$, represents a soft anchoring of $\mathbf{P}$ to the droplet interface (so that $\mathbf{P}$ will tend to point outwards at the droplet perimeter for $\beta > 0$).
{For simplicity this was set to zero in the main text.} Note that for given $V(c)$ and $k$, $\beta$ determines a dimensionless ratio $H = (\gamma_1-\gamma_2)/(\gamma_1+\gamma_2)$ where $\gamma_{1,2}$ are the interfacial tensions with polarity directed along the outward or inward normal. 

\begin{figure}[h]
\begin{centering}
\includegraphics[width=\columnwidth]{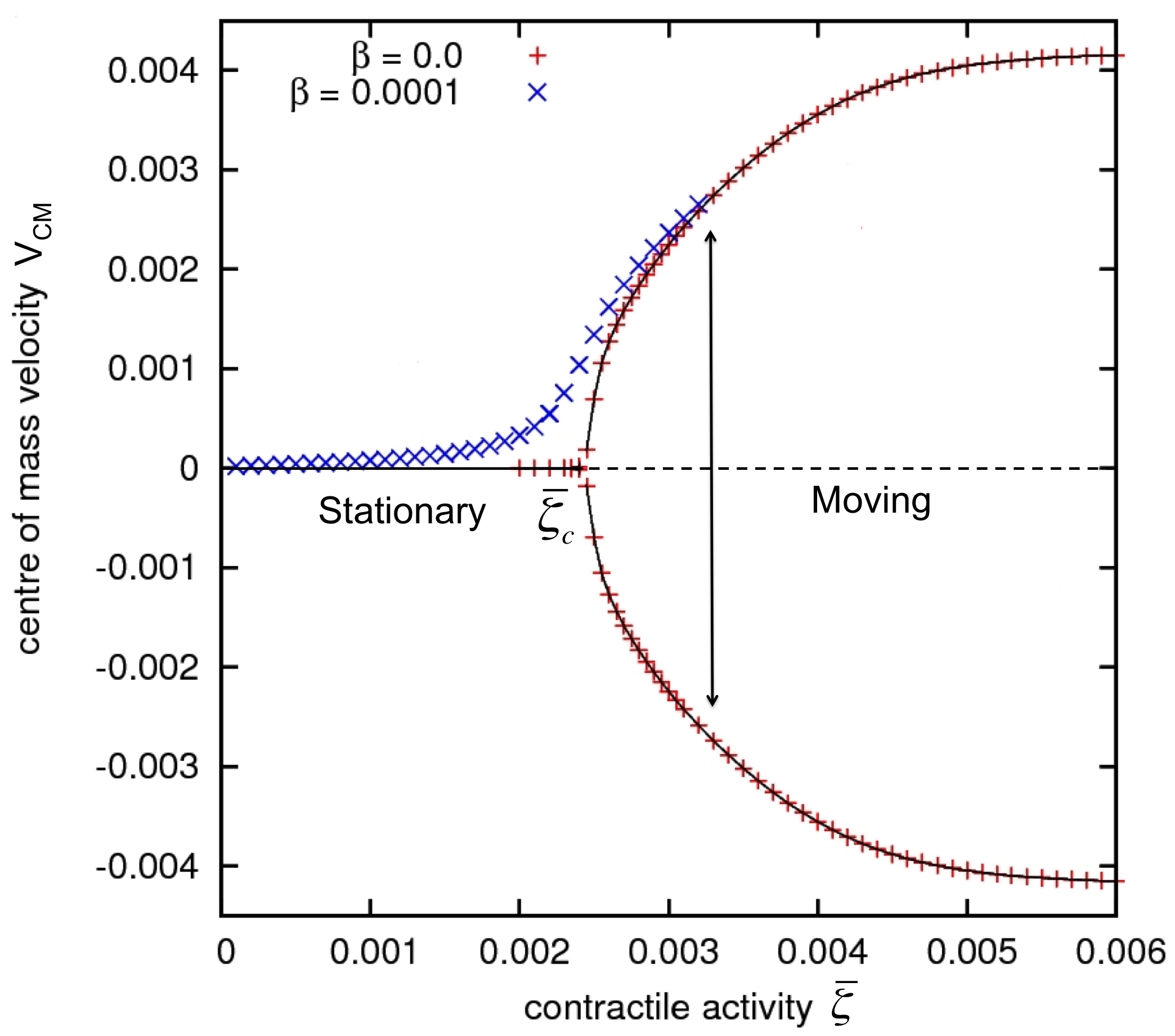}
\end{centering}
\caption{Plots of the droplet velocity as a function of contractile activity for $\beta=0$ 
and $\beta=0.0001$ (roughly corresponds to $H\sim10^{-4}$).
This has a similar effect to a weak external field in para-/ferromagnetic transition which destroys
the second order transition at the critical point. {(Compare Figure 2(c) of main text.)} \label{fig:beta}}
\end{figure}

A nonzero value of $H$ (and hence of $\beta$) is allowed by symmetry and indeed in a passive system of simple amphiphilic molecules one would expect $H$ value of order unity. (The interfacial tension is quite different for amphiphiles correctly oriented at an interface than for ones pointing the wrong way.) However, for an actomyosin droplet, much smaller values of $H$ appear likely: indeed we see no obvious mechanism to suggest a strong preference of either the positive or negative end of an actin fibre for the droplet surface. Accordingly in the main text we set $\beta = 0$, so that the free energy is invariant under the global symmetry transformation of Eq.~\ref{global}. (This can still be broken by $w$.) However, in Fig. \ref{fig:beta} we show the effect of adding a small $\beta$ term on the spontaneous symmetry breaking transition that leads to motility in the absence of self-advection ($w=0$). Here, we choose $\beta=0.0001$ which roughly corresponds to $H\sim10^{-4}$. The effect resembles that of applying a weak external field to a ferromagnet undergoing an Ising-type phase transition. 
Just as in that case, we can expect a detailed understanding of the zero-field ($\beta = 0$) case to offer fruitful mechanistic insights, even if in reality a small field is generically present.

Since the total number of active particles is conserved, the time evolution
of the concentration field $c(\mathbf{r},t)$ can now be written
as a convective-diffusion equation:
\begin{equation}
\frac{\partial c}{\partial t}+\nabla\cdot\left(c(\mathbf{v}+w_1\mathbf{P})-M\nabla\frac{\delta F}{\delta c}\right)=0
\end{equation}
where $M$ is the mobility of the active particles, related to the
diffusion constant by $D\simeq Ma$ and $w_1$ is the self-advection or the
speed of the active particles relative to the bulk fluid. 
{
In the main text, we stated that $w_1$ (there denoted $w$) is related to the velocity of actin treadmilling 
(which is, in turn, proportional to the rate of actin polymerisation at the positive end of the filament).
However, a contribution to $w_1$ may also arise due to other active processes in actomyosin systems such as when the motor proteins preferentially pull the actin filaments towards ther positive ends more favourably. 
Note also that
the ${\delta F}/{\delta c}$ } term is the chemical potential derived
from the free energy in Eq.~\ref{feappendix}.

The dynamics of the polarisation field $\mathbf{P}(\mathbf{r},t)$
is borrowed from polar liquid crystal theory, and can be written
as (see ~\cite{Marchetti} for a more thorough derivation and
justification) 
\begin{equation}
\frac{\partial\mathbf{P}}{\partial t}+\left((\mathbf{v}+w_2\mathbf{P})\cdot\nabla\right)\mathbf{P}=-\underline{\underline{\Omega}}\cdot\mathbf{P}+\xi\underline{\underline{v}}\cdot\mathbf{P}-\frac{1}{\Gamma}\frac{\delta F}{\delta\mathbf{P}}\label{Pdot}
\end{equation}
where $\underline{\underline{v}}$ and $\underline{\underline{\Omega}}$
are the symmetric and anti-symmetric parts of the velocity gradient
tensor $\nabla\mathbf{u}$. $\Gamma$ is the rotational viscosity
and $\xi$ is a shape factor related to the geometry of the active
particles: $\xi>0$ for rod-like particles and $\xi<0$ for oblate
particles. Here we take $\xi$ positive as seems appropriate for filamentary contractile matter such as actin networks. The parameter $\xi$ also 
determines whether the particles
are shear-aligning (for $\left|\xi\right|>1$) or shear-tumbling (for
$\left|\xi\right|<1$). In this paper we assume $\xi>1$. Note that 
Eq.~\ref{Pdot} contains in principle a second self-advection parameter $w_2$ that, for an arbitrary model of activity, need not equal $w_1$. However, it seems highly plausible that (at least for a treadmilling mechanism) both concentration and polarity should advect at the same rate; in this work we therefore set $w_1 = w_2 = w$.  
{Note in addition that an even more complete approach would allow two further terms, $w_3\nabla(P^2)$ and $w_4{\bf P}(\nabla.{\bf P})$ to appear in Eq.~\ref{Pdot} \cite{extra_A}. Inclusion of these terms could lead to a significant rounding of the transition along the lines discussed above for the $\beta$ term.}

Force balance in our system is enforced through the incompressible Navier-Stokes equation, 
\begin{eqnarray}
\nabla\cdot\mathbf{v} & = & 0\\
\rho\left(\frac{\partial}{\partial t}+\mathbf{v}\cdot\nabla\right)\mathbf{v} & = & -\nabla P+\nabla\cdot\underline{\underline{\sigma}}^{total}-\gamma\mathbf{v}
\end{eqnarray}
where $P$ is the isotropic pressure and $-\gamma\mathbf{v}$ is the
friction force, already discussed in the main text. Here $\underline{\underline{\sigma}}^{total}$
is the total hydrodynamic stress which includes the active stress.
There are four contributions to the hydrodynamic stress:
\begin{equation}
\underline{\underline{\sigma}}^{total}=\underline{\underline{\sigma}}^{viscous}+\underline{\underline{\sigma}}^{elastic}+\underline{\underline{\sigma}}^{interface}+\underline{\underline{\sigma}}^{active}
\end{equation}
The first one is the viscous/dissipative stress which can be written
as: $\sigma_{\alpha\beta}^{viscous}=\eta(\partial_{\alpha}v_{\beta}+\partial_{\beta}v_{\alpha})$
where $\eta$ is the shear viscosity and the greek indices indicate
cartesian coordinates. Next there is the elastic/Ericksen
stress borrowed from the liquid crystal dynamics:
\begin{equation}
\sigma_{\alpha\beta}^{elastic}=\frac{1}{2}(P_{\alpha}h_{\beta}-P_{\beta}h_{\alpha})-\frac{\xi}{2}(P_{\alpha}h_{\beta}+P_{\beta}h_{\alpha})-\kappa\partial_{\alpha}P_{\gamma}\partial_{\beta}P_{\gamma}\label{eq:elastic-stress}
\end{equation}
The interfacial stress between the active phase and the passive phase is 
\begin{equation}
\sigma_{\alpha\beta}^{interface}=\left( f-c\frac{\delta F}{\delta c} \right)\delta_{\alpha\beta} - \frac{\partial f}{\partial\left(\partial_{\beta}c\right)} \partial_{\alpha}c
\end{equation}
similar to that of binary fluid~\cite{deGroot-Mazur}, with $f$ defined to be the free energy density. The active stress can be derived
by summing the contributions from each force dipole and coarse-graining~\cite{Ramaswamy} which results in:
\begin{equation}
\sigma_{\alpha\beta}^{active}=\zed cP_{\alpha}P_{\beta}\label{eq:active-stress}
\end{equation}
where $\zed$ is the activity parameter which is positive for contractile
particles (pullers) and negative for extensile particles (pushers).
{Note that active stress contributions proportional to $\partial_{\alpha}P_{\beta}$
are also allowed by symmetry in principle \cite{extra_B}. These higher order terms are plausibly relevant to bacterial swimmers which may have a high degree of asymmetry between the particle's head and tail.
However, for actomyosin solutions (comprising elongated fibres of uniform width, with the activity provided by dilute motor proteins crawling along those fibres) any such terms are likely to be small and we omit them from our model.}  
The magnitude of the activity $\left|\zed\right|$ is proportional
to the strength of the force dipoles. 
The active stress is fundamental in our theory, as it is this term which
drives the system out of equilibrium.

\section{Alternative models}
\label{Qapproach}

As noted previously,  in the absence of self-advection
($w=0$) and with no anchoring term ($\beta=0$), 
the equations above are invariant under the global symmetry of Eq.~\ref{global}. This is spontaneously broken at large enough activity $|\zed|$ to create a new route to motility, as explored in the main text.

As already explained, when defects are allowed for, the breaking of a global $\mathbf{P} \to -\mathbf{P}$ symmetry (which leads to motility in our case) is mathematically distinct from a nematic-to-polar transition. Therefore one possible but inequivalent route to motility, already mentioned in the main text, is where a truly nematic system 
($\underline{\underline{Q}}\neq 0, \mathbf{P} = 0$) becomes polar ($\mathbf{P} \neq 0$). If this happens in a system where treadmilling (say) of individual rods is already present at the molecular level, then this transition should cause the onset of macroscopic self-advection with speed $w\propto|\mathbf{P}|$. 
This describes a mechanistically different way in which
spontaneous symmetry breaking can lead to motility; however no motility would ever arise, by this route, in the absence of self-advection ($w =0$). 

\begin{figure}[h]
\begin{centering}
\includegraphics[width=\columnwidth]{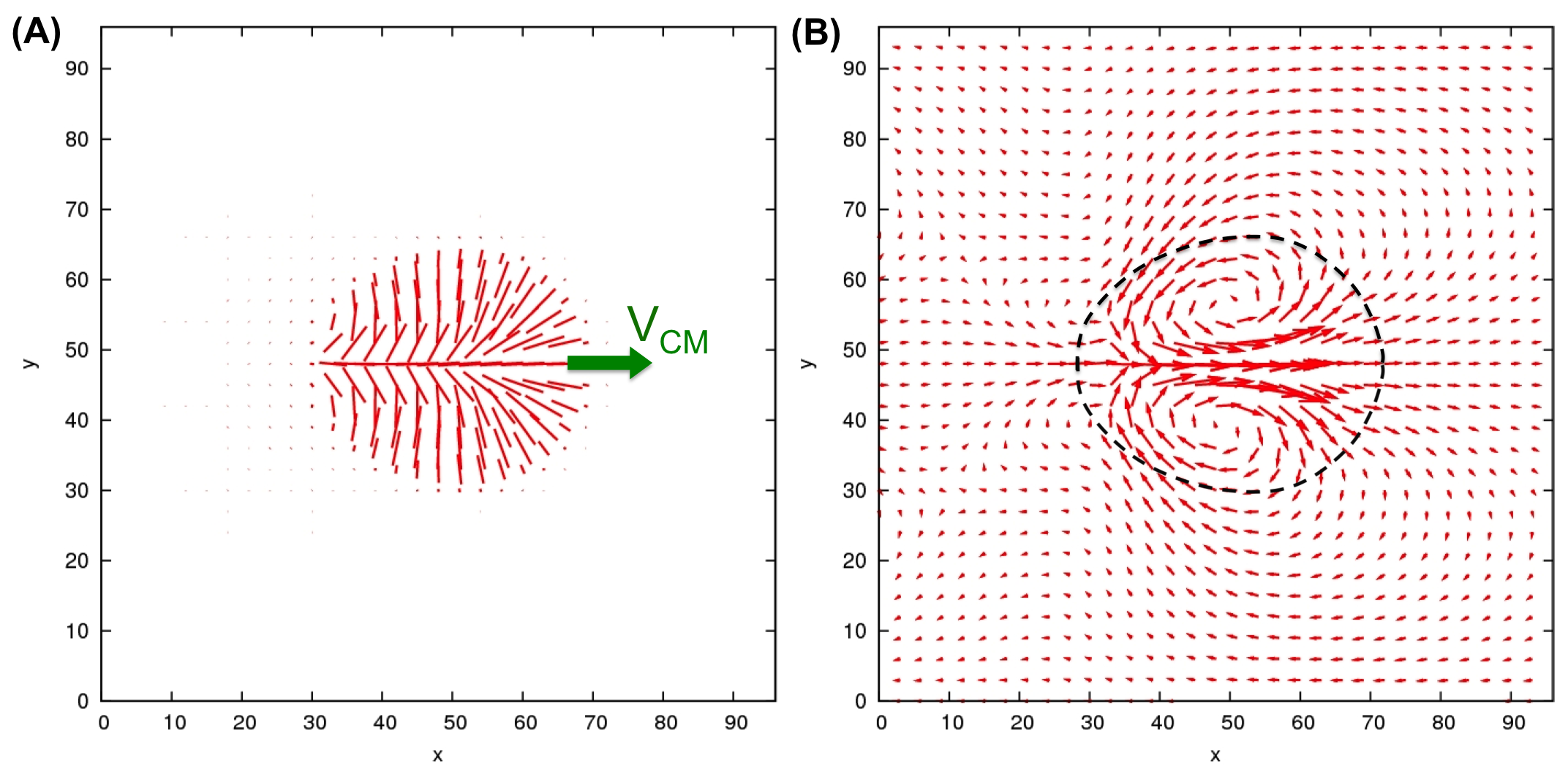}
\end{centering}
\caption{
{A droplet of active nematic ($Q_{\alpha\beta}>0$ and $P_{\alpha}=0$)
can also become motile due to the active stress: $\sigma_{\alpha\beta}=\zed c Q_{\alpha\beta}$. 
\textbf{(A)} shows a typical configuration of the director field in a contractile droplet ($\zed>0$)
moving to the right.
\textbf{(B)} shows the corresponding velocity field consisting of a pair of vortices inside the droplet.
The dashed line represents the droplet interface.}
\label{fig:Q-splay}}
\end{figure}

A way in which spontaneous motility can arise even with $w=0$, which remains mathematically distinct from ours but which is much more closely related to it, is when a truly nematic active droplet ($\underline{\underline{Q}}\neq 0, \mathbf{P} = 0$), spontaneously breaks spatial symmetry to create a splay or bend deformation. For a uniaxial nematic, $Q_{\alpha\beta} = S\left(n_{\alpha}n_{\beta}-\frac{1}{3}\delta_{\alpha\beta}\right)$ with $\mathbf{n}$ the director field; so long as defects are not involved, the dynamics in this case should be extremely similar to a polar system with $\mathbf{P} = S^{1/2}\mathbf{n}$ \cite{Chaikin-Lubensky}. 
{For instance, Fig. \ref{fig:Q-splay} shows one example of a spontaneously moving state in a droplet of an active nematic.
Here, the splay instability due to contractile stress creates a vortex flow inside the droplet which is indeed similar to the polar droplet
case discussed in the main text. 
The droplet shape is somewhat different, but not dissimilar to the ones found for the polar droplet at somewhat higher activity values than shown in Fig. 2(B) of the main text, whose 3D analog is seen in Fig. 4(C).}
 
On the other hand, as shown already in Fig. 4(D) of the main text, defects can and do arise in our active droplet system. If, as we claimed above, nonzero $\mathbf{P}$ is the generic situation for actomyosin gels even under quasi-static conditions and with~$w=0$ \cite{Anderson}, then our description based on the dynamics of $\mathbf{P}$ is preferable on principle to one based on $\underline{\underline{Q}}$. 

\section{Simple scaling analysis}

In the absence of self-advection, we observe a stationary-to-motile
transition at some critical activity $\zed_{c}$ which is accompanied
by spontaneous symmetry breaking of the global symmetry in Eq.~\ref{global}.
As described in the main text, below the critical activity $\zed_{c}$,
the droplet remains stationary while the polarisation field is uniform
inside the droplet. In addition, the droplet is also elongated due
the competition between the active stress and the interfacial tension.
In general the shortest radius of the elongated droplet is a function
of the activity $\zed$ and surface tension $\tilde{\sigma}$, or
$R(\zed,\tilde{\sigma})$.

\begin{figure}[h]
\begin{centering}
\includegraphics[width=\columnwidth]{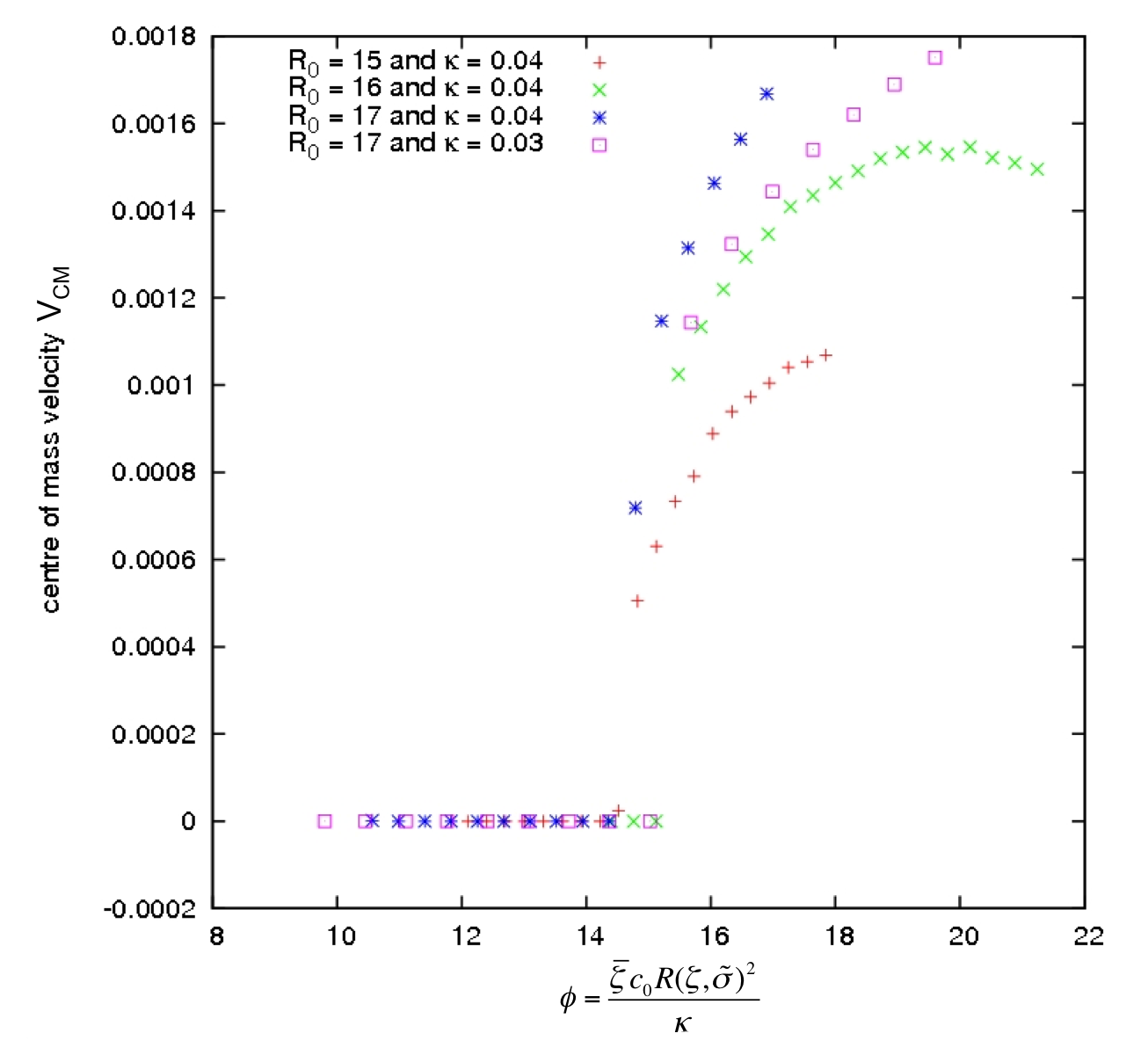}
\end{centering}
\caption{Plots of the droplet speed as a function of dimensionless parameter
$\phi$ for different values of initial droplet radius $R_{0}=R(\zed=0,\tilde{\sigma})$
and elastic constant $\kappa$. \label{fig:scaling}}
\end{figure}

At large enough activity, above the critical value $\zed_{c}$, the
droplet becomes motile. This transition 
may be understood as driven by spontaneous elastic
deformations (bending or splay) -- therefore the critical
value  $\zed_{c}$ can be estimated by
equating the active and elastic stresses, as follows
\begin{equation}
\underline{\underline{\sigma}}^{active}\sim\underline{\underline{\sigma}}^{elastic}.
\end{equation}
Eqs. (\ref{eq:elastic-stress}) and (\ref{eq:active-stress}) lead to
$\underline{\underline{\sigma}}^{active}\sim\zed c_{0}$
and $\underline{\underline{\sigma}}^{elastic}\sim\frac{\kappa}{R^{2}}$,
hence to the following scaling law:
\begin{equation}
\zed_{c}\sim\frac{\kappa}{c_{0}R(\zed_{c},\tilde{\sigma})^{2}}
\end{equation}
which represents a mean-field estimate of the location of the critical point. Notice that the surface tension $\tilde{\sigma}$
will also depend on $\kappa$ in general. 
This scaling law also has the same form as that of 1D spontaneous flow transition~\cite{Voituriez}.
To test this scaling law,
we may plot the droplet speed ${V}_{CM}$ as a function of a dimensionless
quantity:
\begin{equation}
\phi=\frac{\zed c_{0}R(\zed,\tilde{\sigma})^{2}}{\kappa}
\end{equation}
near the critical point for different values of elastic constant $\kappa$ and initial radius of the droplet
$R_0=R(\zed=0,\tilde{\sigma})$.
These plots are given in Fig. \ref{fig:scaling}. We can see from the figure that all the four curves corresponding to different parameters have almost the same critical point at $\phi\simeq14.5$, thereby validating our approximate scaling analysis.

\section{Lattice Boltzmann units}

To establish an approximate correspondence between the natural simulation units (lattice Boltzmann units) and those of typical cell extract experiments, we choose the length-scale, time-scale and force-scale to be: $L=1\,\mu \mathrm{m}$, $T=10\, \mathrm{ms}$, and $F=100\, \mathrm{nN}$ respectively. In lattice Boltzmann/simulation
units these three scales are all defined to be equal to one exactly: that is, the lattice parameter is taken as the unit of length and the timestep the unit of time. 
The variables and parameters used in our model are given in the Table 1
with their respective values in simulation and physical units.
These parameters have been chosen to be roughly in the same order of magnitude to the ones quoted in~\cite{Rubinstein,Theriot-Mogilner}.
Note that, in line with standard practice in Lattice Boltzmann simulations, we chose the fluid mass density $\rho$ to be much larger than the actual mass density of a real solvent (water)~\cite{codef}. This is acceptable so long as the role of inertial (characterized by the Reynolds number) remains small; the procedure speeds up the computations by several orders of magnitude. 
In this case the choice of the force/density scale above gives the Reynolds number of about $\mathrm{Re}\simeq0.03$ which is small enough for the flow to be laminar though much larger than the actual experimental value which is of the order of $\mathrm{Re}\sim10^{-12}$.

\begin{table*}[h]
\raggedright
\caption*{\textbf{Table. 1.} Typical values of the physical quantities used in the simulations.
This choice of parameters are made to be consistent to other physical estimates in~\cite{Rubinstein,Theriot-Mogilner}.
\centering
\label{table}}
\begin{tabular}{c|c|c}
Model variables and parameters & Simulation units & Physical units  \\
\hline 
Effective shear viscosity, $\eta$ 			 & $5/3$ 		 & $1.67\, \mathrm{kPa}/\mathrm{s}$ \\
Effective elastic constant, $\kappa$ 			 & $0.04$ 		 & $4\, \mathrm{nN}$ \\
Shape factor, $\xi$ 					 & $1.1$		 & dimensionless \\
Self-advection/polymerisation speed, $w$	 	 & $0.0015$		 & $0.15\,\mu \mathrm{m}/\mathrm{s}$ \\
Average concentration of actin-myosin pairs, $c_{0}$  	 & $2$ 			 & $2\,\mu \mathrm{m}^{-3}$ \\
Critical concentration, $c_{cr}$	        	 & $1$			 & $1\,\mu \mathrm{m}^{-3}$ \\
Effective diffusion constant, $D=Ma$			 & $0.007$ 		 & $0.7\,\mu \mathrm{m}^{2}/\mathrm{s}$ \\
Rotational viscosity, $\Gamma$ 				 & $1$ 			 & $1\, \mathrm{kPa}\cdot \mathrm{s}$ \\
Friction coefficient (if focal adhesion is present), $\gamma$ & $0-0.04$ & $(0-0.04)\, \mathrm{nN}\cdot \mathrm{s}/\mu \mathrm{m}^{4}$ \\
Activity, $\zed$ 					 & $0-0.01$ 		 & $(0-1)\, \mathrm{nN}\cdot\mu \mathrm{m}$ \\
\end{tabular}
\end{table*}


\begin{thebibliography}{99}

\bibitem{Weijer1} Dormann D., Weijer C.J. (2006) 
Imaging of cell migration. 
{\it EMBO J.} 25:3480-93.
%
\bibitem{Poincloux} Poincloux R., Collin O., Lizarraga F., Romao M., Delbray M., Piel M., Chavrier P. (2011) 
Contractility of the cell rear drives invasion of breast tumor cells in 3D Matrigel. 
{\it Proc. Natl. Acad. Sci. USA} 108:1943-1948.
%
\bibitem{Ratchet} Peskin C. S., Odell G. M., Oster G. F. (1993)
Cellular motions and thermal fluctuations - the Brownian ratchet.
{\it Biophys. J.} 65:316-324.
%
\bibitem{Theriot-Kondev} Phillips R., Kondev J., Theriot J. A. (2008) 
{\it Physical biology of the cells},
(Garland Science).
%
\bibitem{Theriot-Mogilner} Barnhart E. L., Lee K. C., Keren K.,  Mogilner A., Theriot J. A. (2011) 
An adhesion-dependent switch between mechanisms that determine motile cell shape.
{\it PLoS Biology} 9(5):e1001059.
%
\bibitem{Evan-Ram} Evan-Ram S., Yamada K. M. (2005) 
Cell migration in 3D matrix. 
{\it Curr. Opin. Cell Biol.} 17:524-532.
%
\bibitem{Hawkins} Hawkins R. J., Poincloux R., Benichou O., Piel M., Voituriez R. (2011) 
Spontaneous contractility-mediated cortical flows generates cell 
migration in three-dimensional environments. 
{\it Biophys. J.} 101:1041-1045. 
%
\bibitem{Friedl} Friedl P., Wolf K. (2003) 
Tumour-cell invasion and migration: diversity and escape mechanism.
{\it Nat. Rev. Cancer} 3:362-374.
%
\bibitem{Liverpool} Liverpool T. B., Marchetti M. C. (2006)
Rheology of active filament solutions.
{\it Phys. Rev. Lett.} 97:268101.
%
\bibitem{Carlier} Loisel T., Boujemaa R., Pantaloni D., Carlier M.-F. (1999)
Reconstitution of actin-based movement using pure proteins. 
{\it Nature} 401:613-616.
%
\bibitem{Bausch} Köhler S., Schaller V., Bausch A. R. (2011)
Structure formation in active networks. 
{\it Nature Mat.} 10:462–468.
%
\bibitem{Gerisch} Gerisch G., Bretschneider T., Muller-Taubenberger A., Simmeth E., Ecke M. {\it et al.} (2004) 
Mobile actin clusters and traveling waves in cells recovering from actin depolymerization. 
{\it Proc. Natl. Acad. Sci. USA} 87:2493-3503.
%
\bibitem{treadmilling}
Sambeth R., Baumgaertner A. (2001) Autocatalytic polymerization generates persistent random walk of crawling cells. {\it Phys. Rev. Lett.} 86:5196-5199.
%
\bibitem{Wolgemuth} Wolgemuth C. W., Stajic J., Mogilner A. (2011) 
Redundant mechanisms for stable cell locomotion revealed by minimal models. 
{\it Biophys. J.} 101:545-553.
%
\bibitem{Ziebert} Ziebert F., Swaminathan S., Aranson, I. S. (2011)
Model for self-polarisation and motility of keratocyte fragments. 
{\it J. Roy. Soc. Interface}, doi:10.1098.
%
\bibitem{Kruse} Kruse K., Joanny J. F., Julicher F., Prost J., Sekimoto K. (2004) 
Asters,Vortices, and Rotating Spirals in Active Gels of Polar Filaments.
{\it Phys. Rev. Lett.} 92:078101.
%
\bibitem{Tjhung} Tjhung E., Cates M. E., Marenduzzo D. (2011)
Nonequilibrium steady states in polar active fluids. 
{\it Soft Matter} 7:7453-7464. 
%
\bibitem{Theriot} Keren K., Pincus Z. Allen G. M., Barnhart E. L., Marriott G., Mogilner A., Theriot J. A. (2008) 
Mechanism of shape determination in motile cells.
{\it Nature} 453:475-480.
%
\bibitem{Baskaran} 
Baskaran A,. Marchetti M.C. (2009)
Statistical mechanics and hydrodynamics of bacterial suspensions.
{\it Proc. Natl. Acad. Sci. USA} 106:15567-15572.
%
\bibitem{SoftMatter} Cates M. E., Henrich O., Marenduzzo D., Stratford K. (2009)
Lattice Boltzmann simulations of liquid crystalline fluids: active gels and blue phases.
{\it Soft Matter} 5:3791-3800.
%
\bibitem{tumblers} Parkinson J. S., Parker S. R., Talbert P. B, Houts S. E.
(1983) Interactions
between chemotaxis genes and flagellar genes in Escherichia coli. 
{\it J. Bacteriol.} 155:265–274.
%
\bibitem{jana} Schwarz-Linek J., Valeriani C., Cates M.~E., Cacciuto A., Marenduzzo D., Morozov A.~N., Poon W.~C.~K. (2012) 
Phase separation and rotor self-assembly in active particle suspensions.
{\it Proc. Natl. Acad. Sci. USA} 109:4052-4057.
%
\bibitem{Ramaswamy} Hatwalne Y., Ramaswamy S., Rao M., Simha R. A. (2007)
Rheology of active-particle suspensions.
{\it Phys. Rev. Lett.} 92:118101.
%
{
\bibitem{Mitchell} Mitchell P. (1972), Self-electrophoretic locomotion in microorganisms -- bacterial flagella as giant ionophores. {\it FEBS Letters} 28:1.
\bibitem{Lammert} Lammert P. E., Prost J., Bruinsma R. (1996)
Ion drive for vesicles and cells.
{\it J. Theor. Biol.} 178:387-391.
}
%
\bibitem{Simha} 
Simha R. A., Ramaswamy S. (2002) 
Hydrodynamic fluctuations and instabilities in ordered suspensions of self-propelled particles.  
{\it Phys. Rev. Lett.} 89:058101.
%
\bibitem{Giomi-Marchetti} 
Giomi L., Marchetti M. C. (2012) 
Polar patterns in active fluids.
{\it Soft Matter} 8:129-139.
%
\bibitem{harvardSGR} Stamenovic, D., Rosenblatt, N., Montoya-Zavala, M., Matthews, B.D., Hu, S. {\em et al.} (2007) Rheological Behavior of living cells is timescale-dependent. {\it Biophys. J.} 93:L39-L41.
%
\bibitem{Yam} Yam P. T., Wilson C. A., Ji L., Hebert B., Barnhart E. L. {\it et al.} (2007)
Actin-myosin network reorganization breaks symmetry
at the cell rear to spontaneously initiate polarized cell motility.
{\it J. Cell. Biol.} 178: 1207-1221.
%
\bibitem{Chaikin-Lubensky} Chaikin P. M., Lubensky T. C. (2000) 
{\it Principles of condensed matter physics},
(Cambridge University Press).
%
\bibitem{Suzanne} Cates M. E., Fielding S. M., Marenduzzo D., Orlandini E., Yeomans J. M. (2008)
Shearing active gels close to the isotropic-nematic transition.
{\it Phys. Rev. Lett.} 101:068102.
%
\bibitem{Doubrovinski} Doubrovinski K., Kruse K. (2011)
Cell motility resulting from spontaneous polymerization waves.
{\it Phys. Rev. Lett.} 107:258103.

\bibitem{Ranft} Ranft J., Basan M., Elgeti J., Joanny J.-F., Prost J., Julicher F. (2010) 
Fluidization of tissues by cell division and apoptosis. 
{\it Proc. Natl. Acad. Sci. USA} 107:20863-20868.
%
\bibitem{Vasiev}
Vasiev B., Balter A., Chaplain M., Glazier J.~A., Erijer C.~J. (2010)
Modeling gastrulation in the chick embryo: formation of the primitive streak.
{\it PLoS One} 5:e10571.
%
\bibitem{PGG} de Gennes P.-G., Prost. J. (1993) {\it The physics of liquid crystals}, (Clarendon Press, Oxford).
%
\bibitem{Frank} 
Frank F. C. (1958)
On the theory of liquid crystals.
{\it Discussions Faraday Soc.} 25:19.

\bibitem{Ericksen} 
Ericksen J. L. (1961)
Conservation laws for liquid crystals.
{\it Trans. Soc. Rheol.} 5:23.

\bibitem{Yeomans} 
Edwards S. A., Yeomans J. M. (2009)
Spontaneous flow states in active nematics: a unified picture
{\it EPL} 85:18008.

\bibitem{deGroot-Mazur} deGroot S. R., Mazur P. (2011) 
{\it Non-Equilibrium Thermodynamics},
(Dover Publications).

\bibitem{Anderson} Small J. V., Herzog M., Anderson K. (2008)
Actin filament organization in the fish keratocyte lamellipodium.
{\it J. Cell Biol.} 129:1275-1286.

\bibitem{Marchetti} Kung W., Marchetti M. C., Saunders K. (2006)
Hydrodynamics of polar liquid crystals.
{\it Phys. Rev. E} 73:031708.

\bibitem{extra_A} Bertin E., Droz M., Gregoire G. (2006)
Boltzmann and hydrodynamic descriptions for self-propelling particles.
{\it Phys. Rev. E} 74:022101. 

\bibitem{extra_B} Marchetti M. C., Liverpool T. B. (2007) 
Hydrodynamics and rheology of active polar filaments, in 
{\it Cell Motility}, ed. P. Lenz, Springer-Verlag, NY.

\bibitem{Voituriez} Voituriez R., Joanny J. F., Prost J. (2005)
Spontaneous flow transition in active polar gels.
{\it EPL} 70(3):404-410.

\bibitem{codef} Cates M. E., Stratford K., Adhikari R., Stansell P., Desplat J.-C., Pagonabarraga I., Wagner A. J. (2004)
Simulating colloid hydrodynamics with lattice Boltzmann methods.
{\it J. Phys. Cond. Mat.} 16:S3903-S3915.

\bibitem{Rubinstein} Rubinstein B., Fournier M. F., Jacobson K., Verkhovsky A., Mogilner A. (2009)
Actin-myosin viscoelastic flow in the keratocyte lamellipod.
{\it Biophys.} 97:1853-1863.

\end{thebibliography}
\end{document}